# Pressure and strain effects on the optical properties of $K_4$ phosphorus

Qun Wei [a,*], Chenyang Zhao [b], Meiguang Zhang [c], Haiyan Yan [d], Quan Zhang [b], Yingjiao Zhou [a], Xuanmin Zhu [a]

[a] School of Physics and Optoelectronic Engineering, Xidian University, Xi'an 710071, PR China
[b] School of Microelectronics, Xidian University, Xi'an 710071, PR China
[c] College of Physics and Optoelectronic Technology, Baoji University of Arts and Science, Baoji 721016, PR China
[d] College of Chemistry and Chemical Engineering, Baoji University of Arts and Sciences, Baoji 721013, PR China

**ABSTRACT**

An investigation of the mechanical, electronic, and optical properties of the recently reported material $K_4$ phosphorus was made in this work. The $K_4$ phosphorus has been proved to be mechanically and dynamically stable up to 7 GPa under hydrostatic pressure. We compared the elastic anisotropy, average acoustic velocity and Debye temperature of $K_4$ phosphorus at 0 and 7 GPa. The ideal tensile at large strains of $K_4$ phosphorus was also examined, with the results showing that it would cleave under the tensile strength of 8.5 GPa with the strain of 0.3. In addition, the effect of tensile strain and pressure on optical properties and band gap were studied.

# 1. Introduction

In industry and materials science, phosphorus has been a constant source of concern for researchers [1]. As known, at ambient conditions, white phosphorus, orthorhombic black phosphorus, and various forms of red phosphorus are the stable allotropes of phosphorus. Researchers have never stopped working on phosphorus, and a many allotropes of phosphorus have been synthesized, meanwhile, there are still many allotropes of phosphorus that are in the prediction phase [2, 3]. 2D few-layer black phosphorus [4-6] have been successfully fabricated, and researchers have found that this material is chemically inert and has remarkable transport properties. It was reported that 2D few-layer black phosphorus has a carrier mobility up to 1000 $cm^2/V$ s and an on/off ratio up to $10^4$ was achieved for the phosphorene transistors at room temperature [4, 5]. Researchers have been exploring its structural features [7], electronic properties [8], unique mechanical features [9], and so on. The researchers' quest for a new phase of phosphorus has never stopped [10, 11]. After the identification of $K_4$ geometry in mathematics [12], researchers proposed a boron $K_4$ crystal that is stable under ambient pressure [13]. Moreover, under the pressure above 110 GPa, researchers have successfully synthesized the nitrogen $K_4$ crystal from molecular nitrogen using a laser-heated diamond cell [14]. As known, nitrogen, as the element of the fifth main group of the periodic table, is similar in some respects to phosphorus. Recently, inspired by the unique geometry of $K_4$ nitrogen structure, researchers proposed a $K_4$ phosphorus structure [15], which is a semiconductor with indirect band gap and many unique properties. Pressure usually has a great impact on the structural properties of the semiconductor [16, 17]. Electron characteristics are a very important property of semiconductor nanostructures, and strain has been a very common method to regulate the electronic properties of materials [18, 19]. The study found that nanostructures can still maintain their integrity under great strain [20, 21], which is of great significance to the strength of the extended strain to nanostructures [22-24]. These studies motivate us to study the pressure and strain effects on the properties of $K_4$ phosphorus. In this work, at first, we checked the stability of $K_4$ phosphorus under a hydrostatic pressure. Above 7 GPa, the mechanical stability criteria cannot be satisfied. We calculated the band structure and density of states (DOS) of $K_4$ phosphorus at 0 GPa and compared the elastic anisotropies at 0 GPa and 7 GPa. We calculated the ideal tensile stress-strain curves to check the stable range under tensile strain, and studied the influence of tensile strain and pressure on the optical properties and band gap of $K_4$ phosphorus.

## 2. Computational methods

Our calculations were performed based on the density functional theory (DFT) [25, 26] as implement in Cambridge Serial Total Energy Package (CASTEP) code [27]. We employed the Vanderbilt ultrasoft pseudopotentials to describe the electronic interactions in the calculations. The exchange correlation energy was described in the generalized gradient approximation (GGA) using the Perdew-Burke-Enzerhof (PBE) functional [28]. The equilibrium crystal structures were achieved by utilizing geometry optimization in the Broydn-Fletcher-Goldfarb-Shanno (BFGS) [29]. In our calculations, the electronic wave functions were extended in a plane-wave basis set with energy cutoff of 440 eV. The special k-point method proposed by Monkhorst-Pack [30] was used to character energy integration in the first irreducible Brillouin zone, the *k*-point mesh was taken as $10\times10\times10$. Bulk modulus, shear modulus, Young's modulus, and Poisson's ratio were estimated by using Voigt-Reuss-Hill approximation [31]. In addition, we re-calculated the elastic constants calculations by using the Vienna Ab-initio Simulation Package (VASP) [32] at GGA-PBE level. The electron-ion interaction was described by the frozen-core all-electron projector augmented wave (PAW) method [33]. The energy cutoff of 500 eV and a $10\times10\times10$ *k*-point grid were used in the calculations. The Heyd-Scuseria-Ernzerhof (HSE06) hybrid functional [34] was used for the high accuracy of electronic structure calculations and the optical properties calculations.

## 3. Results and discussion

$K_4$ phosphorus has an $I2_13$ symmetry and belongs to the space group No. 199, we have established the structure of $K_4$ phosphorus and optimized the lattice parameters in conventional cell and optimized the cell, as listed in Table 1. The optimized structures of $K_4$ phosphorus at 0 GPa are shown in Fig. 1. We can see that, the lattice parameters we calculated are in an excellent agreement with previous results, so our calculations are reliable. The optimized structure of $K_4$ phosphorus is body-centered cubic, and in the conventional cell of $K_4$ phosphorus, there are 8 atoms located at the 8a (0.204, 0.204, 0.204) Wyckoff position.

The central phosphorus atom is $sp^3$-hybridized, which is connected to its three neighboring atoms with equal bond lengths and band angles. The bond lengths at 0 GPa become 2.23 Å with the bond angles of 102.3°, as the pressure increases to 7 GPa, they are 2.21 Å with the bond angles of 99.5°. Both the bond lengths and bond angles decrease when the pressure increases.

Elastic properties have always been a very important part of the many properties of materials. According to the elastic constant, we can obtain the stability, hardness, and anisotropy information of the materials. In a cubic phase structure, the mechanical stability criterions for $K_4$ phosphorus at 0 GPa are [35]:

$$C_{11} > 0, C_{44} > 0,\ C_{11} > |C_{12}|,\ (C_{11} + 2C_{12}) > 0 . \tag{1}$$

Under isotropic pressure, the criteria of mechanical stability are provided by Ref. [36]:

$$\tilde{C}_{11} > 0, \tilde{C}_{44} > 0,\ \tilde{C}_{11} > |\tilde{C}_{12}|,\ (\tilde{C}_{11} + 2\tilde{C}_{12}) > 0 , \tag{2}$$

where

$$\tilde{C}_{11} = C_{11} - P, \tilde{C}_{44} = C_{44} - P, \tilde{C}_{12} = C_{12} + P , \tag{3}$$

where $P$ is the isotropic pressure.

The calculated crystal elastic constants $C_{ij}$ of $K_4$ phosphorus under various pressures are shown in Table 2. We can see that the calculated elastic constants at 0 GPa are much smaller than those in Ref. [15]. Then we re-calculated the elastic constants by using the strain-stress method [37] which is implemented in VASP code, the results are also listed in Table 2. Clearly, the results calculated by using VASP are in a good agreement with those calculated via CASTEP. So, our results are reliable. From Table 2, we can also see that the $K_4$ phosphorus is mechanically stable up to 7 GPa. In addition, to ensure its dynamical stability at 7 GPa, we calculated the phonon spectra. As Fig. 2 (a) shows, there is no imaginary frequency in the whole Brillouin zone, indicating that $K_4$ phosphorus is dynamically stable up to 7 GPa.

In order to better characterize the intrinsic hardness of $K_4$ phosphorus, the ideal stress-strain curves for large strains were calculated. The ideal strength of a material generally means that the stress which is applied to a perfect crystal and make the crystal mechanically unstable. This stress sets an upper limit for material strength. For cubic phase, we only need to consider the $a$-direction as the tension direction. As seen from Fig. 2 (b), the ideal tensile strength is 8.5 GPa with the strain of 0.3, which means that the $K_4$ phosphorus would cleave if the strain is larger than 0.3.

In crystal physics and engineering science, the bulk modulus $B$ denotes the resistance to fracture, and the shear modulus $G$ represents the resistance to plastic deformation. According to the Voigt-Reuss-Hill approximations [31, 38, 39]:

$$B_V = B_R = (C_{11} + 2C_{12})/3, \tag{4}$$

$$G_V = (C_{11} - C_{12} + 3C_{44})/5, \tag{5}$$

$$G_R = 5(C_{11} - C_{12})C_{44} / [4C_{44} + 3(C_{11} - C_{12})], \tag{6}$$

we can obtain the bulk modulus $B$ and shear modulus $G$:

$$B = \frac{1}{2}(B_V + B_R), \tag{7}$$

$$G = \frac{1}{2}(G_V + G_R). \tag{8}$$

The ratio of bulk to shear modulus ($B/G$) proposed by Pugh [40] is an indication of ductile or brittle character. A larger $B/G$ ratio is associated with a more facile ductility, whereas a smaller $B/G$ ratio corresponds to a brittle nature. If $B/G > 1.75$, we consider that the material is ductile [41]; otherwise, the material is brittle. From Table 2, we can see that the $B/G$ ratio of $K_4$ phosphorus is 1.44 at 0 GPa and 2.37 at 7 GPa, thereby, as pressure increases from 0 GPa to 7 GPa, $K_4$ phosphorus changes from being brittle to being ductile.

To get more information about the elastic properties, Young's modulus $E$ and Poisson's ratio $v$ were calculated. As a measure of the stiffness of a solid material, Young's modulus $E$ is defined as the ratio between stress and strain. When a material receives a tension or compression in unidirectional, the absolute value of the ratio of transverse contraction strain to longitudinal extension strain is called Poisson's ratio $v$. They are given by [31, 42]:

$$E = \frac{9BG}{3B + G}, \tag{9}$$

$$v = \frac{3B - 2G}{2(3B + G)}. \tag{10}$$

From Table 2, the Young's modulus $E$ of $K_4$ phosphorus at 7 GPa is smaller than that at 0 GPa. As known, the larger the value of $E$, the stiffer the material. We can see that at 0 GPa, $K_4$ phosphorus possesses the largest stiffness, and when the pressure increases, the stiffness gets smaller. The typical value of $v$ is 0.1 for covalent materials and 0.33 for metallic materials [43]. We can see that the Poisson's ratios of $K_4$ phosphorus increases when pressure increases, indicating that the directionality degree of covalent bonding becomes weaker when pressure increases.

It is very important to calculate the elastic anisotropy of crystal for the study of its physical and chemical properties. For cubic symmetry, we follow Zener [44] and use $A = 2C_{44} / (C_{11}-C_{12})$ to calculate the elastic anisotropy. The value of 1.0 indicates isotropy, and any deviation from 1.0 indicates a degree of the shear anisotropy. The calculated Zener anisotropy factor $A$ at 0 and 7 GPa are listed in Table 2, which show an anisotropy that increase with pressure. To show the elastic anisotropy of $K_4$ phosphorus at 0 and 7 GPa in detail, Young's modulus for all possible directions are shown in Fig. 3. For an isotropic system, the 3D directional dependence shows a spherical shape, while the deviation degree from the spherical shape represents the anisotropy [45]. At 0 GPa, the maximum of Young's modulus $E_{max}$ is 120 GPa, while the minimum of Young's modulus $E_{min}$ is 54 GPa and the average value over all directions is 80 GPa. The ratio $E_{max}/E_{min}$ is 2.22. At 7 GPa, the maximum of Young's modulus $E_{max}$ is 191 GPa, while the minimum of Young's modulus $E_{min}$ is 23 GPa and the average value over all directions is 62 GPa, with a ratio $E_{max}/E_{min}$ of 8.30. These results indicate that the anisotropy increases when the pressure increases, which is consistent with what we have found before.

The acoustic velocity is of great significance to the study of the chemical bonding characteristics, and the symmetry and propagation direction of the crystal determine the acoustic velocity. According to the single crystal elastic constants, Brugger [46] has successfully calculated the phase velocities of pure transverse and longitudinal modes. The cubic structure has three direction [001], [110], and [111] for the pure transverse and longitudinal modes and other directions are for the quasi-transverse and quasi-longitudinal waves. For a cubic phase, the acoustic velocities in the principal directions are [47]:

for [100],

$$v_l = \sqrt{C_{11}/\rho}, \quad [010]v_{t1} = [001]v_{t2} = \sqrt{C_{44}/\rho}, \tag{11}$$

for [110],

$$v_l = \sqrt{(C_{11} + C_{12} + 2C_{44})/2\rho}, \quad [1\bar{1}0]v_{t1} = \sqrt{(C_{11} - C_{12})/2\rho}, \quad [001]v_{t2} = \sqrt{C_{44}/\rho}, \quad (12)$$

for [111],

$$v_l = \sqrt{(C_{11} + 2C_{12} + 4C_{44})/3\rho}, \quad [11\bar{2}]v_{t1} = v_{t2} = \sqrt{(C_{11} - C_{12} + C_{44})/3\rho}, \quad (13)$$

where $v_l$ is the longitudinal acoustic velocity, $v_{t1}$ and $v_{t2}$ are the first transverse mode and the second transverse mode, respectively. The density of the structure is $\rho$. The acoustic velocities are calculated based on the elastic constants. Therefore, we can see the anisotropy of elasticity according to the anisotropy of the acoustic velocities. Calculated the average longitudinal acoustic velocity is $v_{lm} = \sqrt{(B + 4G/3)/\rho}$, and the average transverse acoustic velocity is $v_{tm} = \sqrt{G/\rho}$. The average acoustic velocity is $v_m = \left[(2/v_{tm}^3 + 1/v_{lm}^3)/3\right]^{-1/3}$.

Based on the calculation results of the average acoustic velocity, we can get the Debye temperature:

$$\Theta_D = \frac{h}{k_B}\left[\frac{3n}{4\pi}\left(\frac{N_A \rho}{M}\right)\right]^{1/3} v_m, \quad (14)$$

where $N_A$ is Avogadro's number, $h$ and $k_B$ are the Planck and Boltzmann constants, respectively, $n$ is the total number of atoms in the formula unit, $M$ is the mean molecular weight, and $\rho$ is the density.

The acoustic velocities and Debye temperatures of $K_4$ phosphorus at 0 and 7 GPa are listed in Table 3. We can see that the densities increases when the pressure increases. As the pressure increases from 0 GPa to 7 GPa, the average acoustic velocity decreases by 11.36%, and the Debye temperature decreases by 7.34%. The strength of the covalent bond in solids is related to Debye temperature, so the strength of the covalent bond becomes weaker for $K_4$ phosphorus when the pressure increases.

Electronic structure is also crucial for studying the physical and chemical properties of materials. Thus, we calculated the band structure and density of state (DOS) of $K_4$ phosphorus at 0 GPa. Calculated electronic band structure of $K_4$ phosphorus in primitive cell is plotted in Fig. 4 (a). The black dashed line represents the Fermi level ($E_F$). The black curve represents the results we calculated using the PBE functional. The black arrow points from the valence band maximum (VBM) to the conduction band minimum (CBM). The VBM locates at

(0.3077, -0.3077, 0.3077) along the G-H direction, the CBM locates at (0.0385, -0.0385, 0.0385) along the G-H direction, and the band gap is 1.08 eV. Clearly, the $K_4$ phosphorus is an indirect semiconductor. It is known that the band gap calculated by DFT usually should be smaller than the real values, thus we use the more precise HSE06 functional to correct the band gap of $K_4$ phosphorus. As shown in Fig. 4(a), the red curve represents the results we calculated using HSE06 functional. It can be seen that, using both functionals, similar band structures were obtained. The VBM and CBM calculated by HSE06 functional locate at the same positions as those calculated by PBE functional, and the band gap of $K_4$ phosphorus calculated with HSE06 functional increases to 1.74 eV.

The DOS of $K_4$ phosphorus in a single atom is shown in Fig. 4(b). The black dashed line represents the Fermi level ($E_F$). As Fig. 4(b) shows, we can see that the valence band region can be divided into two parts, the first part (-15 to 7 eV) is characterized by the contributions of $s$ states, the second part (-7 to 0 eV) originates from the contributions of $p$ states. The conduction band region is mainly characterized by the $p$ states. The DOS near Fermi level are mainly originated from the $p$ orbital electrons.

To study the influence of hydrostatic pressure on the energy band gap of $K_4$ phosphorus, we calculated the band gap as a function of pressure. Usually, the trends of change of the band gap calculated by PBE functional and HSE06 functional are similar [48], thus we use PBE functional to calculate the band gap. The changes of band gap are shown in Fig. 5, and we found that with the increase of pressure, the band gap is diminishing, and the band gap decreases almost linearly from 1.08 eV to 0.33 eV when the pressure increases from zero to 7 GPa. We also calculated the band gap as a function of strain. As Fig. 5 shows, when the $K_4$ phosphorus is stretched from the strain of 0.0 to 0.1, the band gap decreases almost linearly from 1.08 eV to 0.79 eV. When it is contracted from the strain of 0.0 to -0.1, the band gap decreases from 1.08 eV to 0.21 eV. Through the above analyses, we conclude that irrespective of whether the $K_4$ phosphorus is pressurized or strained, the band gap will get smaller. As we know, the value of band gap is crucial for efficient optoelectronic devices. Thereby, we can artificially control the strain and pressure to choose any desired value.

We calculated the imaginary parts of dielectric function ($\varepsilon_2$) to study the influences of pressure and strain on the optical absorption properties. The spectral range is divided into three domains which are infrared, visible, and ultraviolet regions, from left to right. As Fig. 6 (a) shows, in the visible region, the optical absorption of $K_4$ phosphorus is much stronger than that of diamond silicon, and with the pressure increase on the $K_4$ phosphorus, the absorption coefficients have a great improvement, and the absorption coefficients of $K_4$ phosphorus are getting closer to the solar spectrum. When the pressure reaches 7 GPa, $K_4$ phosphorus has the best optical absorption properties. In ultraviolet regions, with the pressure increase, the absorption coefficients have a small improvement.

The strain effect on the absorption coefficients is also obvious. As shown in Fig. 6 (b) for the visible region, when the strain applied to $K_4$ phosphorus is from 0 to -0.08, the absorption coefficients have a great improvement, and the optical absorption is stronger than that of diamond silicon. Meanwhile, the absorption coefficients of $K_4$ phosphorus are getting closer to the solar spectrum. When the strain applied to $K_4$ phosphorus is -0.08, they have the best optical absorption properties. In the ultraviolet region, when the strain applied to $K_4$ phosphorus is from 0 to -0.08, the absorption coefficients also have a great improvement, but when the energy is greater than 4 eV, $K_4$ phosphorus exhibits worse optical absorption than diamond silicon. When the strain applied to $K_4$ phosphorus is from 0 to 0.08, the changes of absorption coefficients are tiny, and the optical absorption properties are get a little worse. All the strong adsorption coefficients are in the visible range of sunlight. Therefore, in the study of the absorption characteristics of the material in the visible region, $K_4$ phosphorus would play a crucial role in its promotion.

4. **Conclusions**

In summary, we confirmed the mechanical and dynamic stabilities of $K_4$ phosphorus up to 7 GPa under a hydrostatic pressure. We have calculated the lattice parameters, cell volume, elastic constants, bulk modulus, shear modulus, Young's modulus Poisson's ratio, average acoustic velocity, and the Debye temperature of $K_4$ phosphorus at 0 and 7 GPa, and found that with the pressure increasing, the stiffness get smaller, the anisotropy increases, and the average acoustic velocity and the Debye temperature decreases. The ideal tensile strength is 8.5 GPa and the strain is 0.3. By studying the band gap of $K_4$ phosphorus as a function of hydrostatic pressure and strain, we find that the pressure and strain will all reduce the band gap. We also calculated the imaginary parts of dielectric function to study the influences of pressure and strain on the optical absorption properties, from which we find that when the pressure increase, the optical absorption properties get improved, and when the strain applied to v phosphorus is from 0 to -0.08, the optical absorption properties also improve, but when the strain is from 0 to 0.08, the optical absorption properties become worse.

**Acknowledgments**

This work was supported by the Natural Science Foundation of China (Grant No. 11204007), Natural Science Basic Research plan in Shaanxi Province of China (Grant No.: 2016JM1026, 2016JM1016), the 111 Project (B17035), and Education Committee Natural Science Foundation in Shaanxi Province of China (Grant No.: 16JK1049). Xiao-Feng Shi is acknowledged for the helpful discussions and comment on the manuscript.

**Figures captions:**

**Fig. 1.** The perspective view of the conventional unit cell (a) and crystal structure viewed from the [001] direction (b) of $K_4$ phosphorus at 0 GPa.

**Fig. 2.** The phonon spectra of $K_4$ phosphorus at 7 GPa (a), and the tensile strength of $K_4$ phosphorus (b).

**Fig. 3.** The direction dependence of Young's modulus (GPa) for $K_4$ phosphorus at 0 GPa (a) and 7 GPa (b).

**Fig. 4.** The band structure (a) and density of states (b) of $K_4$ phosphorus.

**Fig. 5.** Variation of band gap versus pressure and strain of $K_4$ phosphorus

**Fig. 6.** Calculated imaginary parts of the dielectric function ($\varepsilon 2$) as a function of energy under pressures (a) and strains (b).

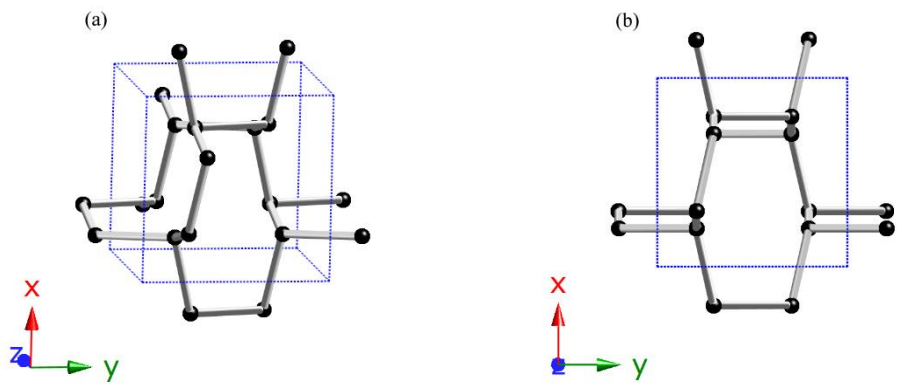

**Fig. 1.**

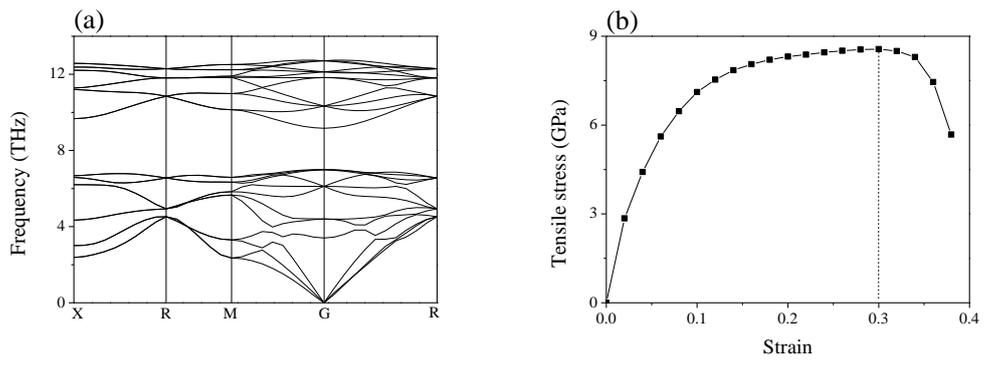

**Fig. 2.**

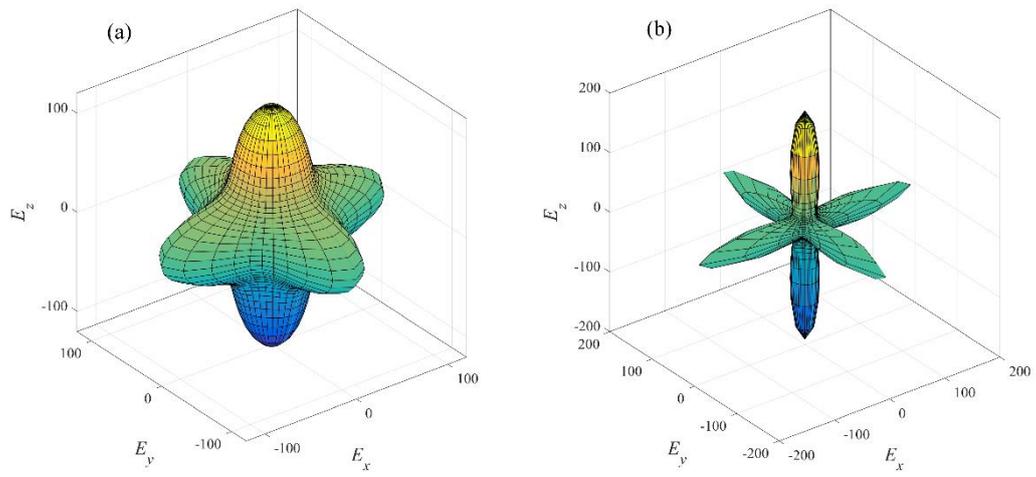

**Fig. 3.**

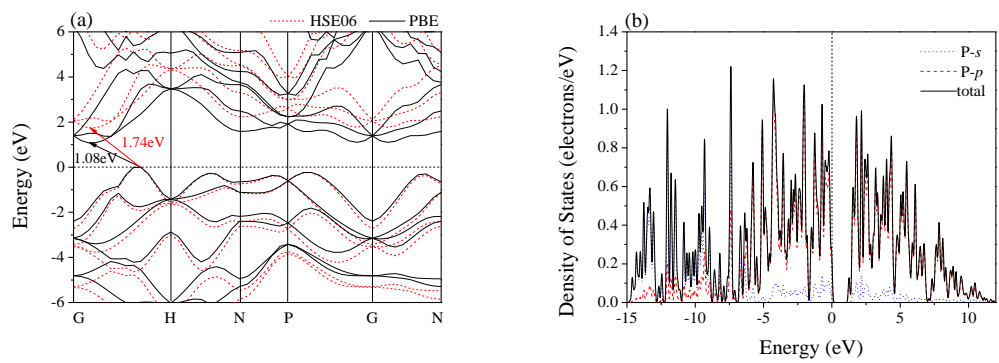

**Fig. 4.**

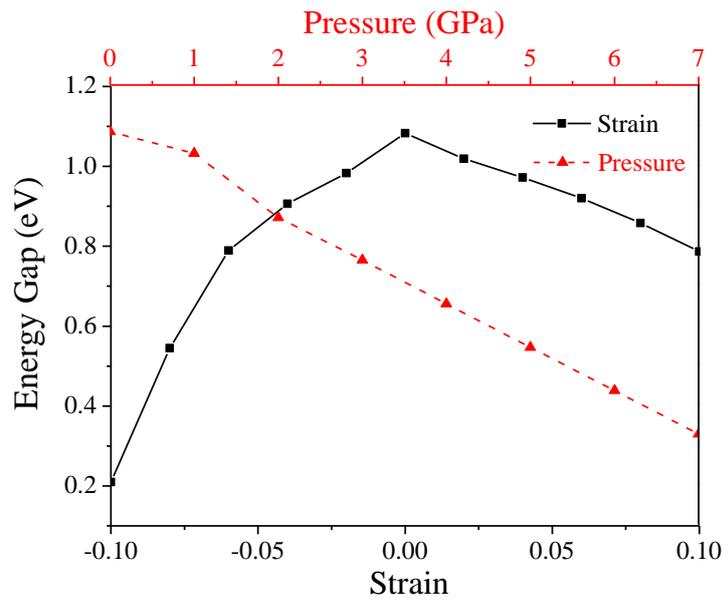

**Fig. 5.**

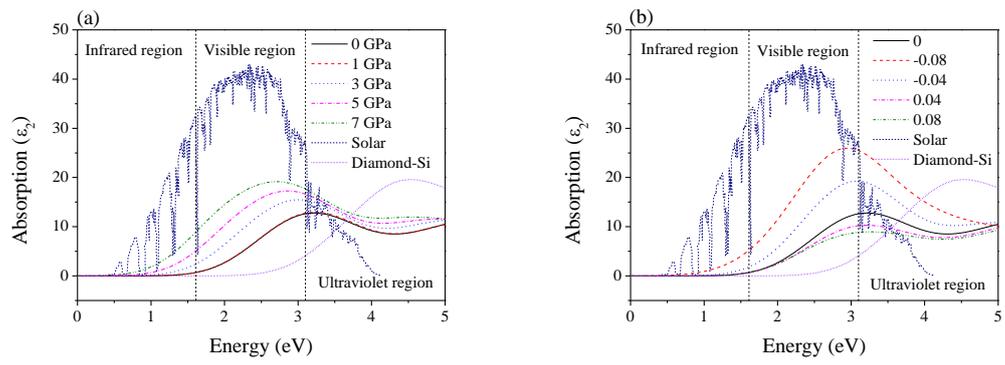

**Fig. 6.**

**Table 1.** Calculated lattice parameters (in Å), band length $d$ (in Å), band angle $\beta$, and volume $V_0$ (in Å$^3$)..

| Pressure (GPa) | $a$ | $d$ | $\beta$ | $V_0$ |
|---|---|---|---|---|
|   | 5.3272 | 2.23 | 102.3 | 151.18 |
| 0 | 5.32[a] |  | 101.7[a] | 150.57[a] |
|   | 5.37[b] |  | 101.7[b] | 154.85[b] |
| 7 | 5.0968 | 2.21 | 99.5 | 132.40 |

[a] Previous calculated results at GGA-D2 level in Ref. [15].
[b] Previous calculated results at PBE level in Ref. [15].

**Table 2.** Calculated elastic constants $C_{ij}$ (GPa), bulk modulus $B$ (GPa), shear modulus $G$ (GPa), Young's modulus $E$ (GPa), Poisson's ratio $v$, $B/G$ ratio, and the Zener anisotropy factor $A$ of $K_4$ phosphorus..

| P(GPa) | $C_{11}$ | $C_{44}$ | $C_{12}$ | $B$ | $G$ | $E$ | $v$ | $B/G$ | $A$ |
|---|---|---|---|---|---|---|---|---|---|
| 0 | 121.5[a] | 21.0[a] | 7.8[a] | 45.7[a] | 31.7[a] | 77.24[a] | 0.22[a] | 1.44[a] | 0.37[a] |
|   | 129.2[b] | 14.9[a] | 1.8[a] |         |         |          |         |         |         |
|   | 223.6[c] | 34.6[b] | 82.4[b] | 129.5[b] |        |          |         |         |         |
|   | 232.2[d] | 31.7[c] | 39.3[c] | 103.6[c] |        |          |         |         |         |
| 7 | 191.9 | 8.2 | 3.0 | 66.0 | 27.8 | 73.1 | 0.32 | 2.37 | 0.09 |
| 8 | 194.6 | 5.8 | 6.5 |  |  |  |  |  |  |

[a] Our results calculated by using CASTEP.

[b] Our results calculated via VASP.

[c] Previous calculated results at PBE level, Ref [15].

[d] Previous calculated results at GGA-D2 level, Ref [15].